\documentstyle[aps,prd,epsf,eqsecnum]{revtex}

\def\e{{\,\rm e}\,}
\def\const{{\rm const}}
\def\A{{\cal A}}
\def\F{{\cal F}}
\def\L{{\cal L}}
\def\bx{{\bf x}}
\def\by{{\bf y}}
\def\bz{{\bf z}}
\def\bX{{\bf X}}
\newcommand{\rf}[1]{(\ref{#1})}
\newcommand{\non}{\nonumber \\*}
\newcommand{\no}[1]{\,:\!#1\!:\,}

\begin{document}

\draft
\preprint{ITEP--TH--38/97}

\title{ Phase structure and nonperturbative states
\protect\\
 in three--dimensional adjoint Higgs model }

\author{N.O. Agasyan\thanks{e-mail: agasyan@vxitep.itep.ru}
and K. Zarembo\thanks{e-mail: zarembo@vxitep.itep.ru}}

\address{
{\it  Institute of Theoretical and Experimental Physics,
B. Cheremushkinskaya 25, 117259 Moscow, Russia}}

\maketitle
\begin{abstract}
The thermodynamics of 3d adjoint Higgs model is considered. We study
the properties of the Polyakov loop correlators and the critical
behavior at the deconfinement phase transition. Our main tool is a
reduction to the 2d sine--Gordon model. The Polyakov loops appear to
be connected with the soliton operators in it. The known exact
results in the sine--Gordon theory allow us to study in detail the
temperature dependence of the string tension, as well as to get some
information about a nonperturbative dynamics in the confinement
phase. We also consider the symmetry restoration at high temperature
which makes it possible to construct the phase diagram of the model
completely.
\end{abstract}


\section{Introduction}

 The adjoint Higgs model in three dimensions exhibits a number of
 features which probably are shared by four dimensional
 gauge theories in the confining phase.
 This theory possesses a mass gap, although a part of the gauge
 symmetry remains unbroken.
 The charges of the unbroken subgroup are
 confined by a string of electric flux with the energy proportional to its
 length.
 The confinement, as well as the mass gap,
 arise nonperturbatively due to the Euclidean field configurations of the
 magnetic monopole type \cite{Polyakov}.

 In three dimensions, the magnetic charge is a counterpart of the
 instanton number. The classical solutions which carry a unit of
 the magnetic charge are the well known 't~Hooft--Polyakov
 monopoles \cite{TP}. The effects driven by these pseudoparticles can be
 studied at weak coupling by standard semiclassical techniques and the
 nonperturbative phenomena can be investigated in much detail, without
 any uncontrollable approximations.

 At low energies, the relevant degrees of freedom in this model are
 gauge fields of the unbroken Abelian subgroup and monopoles.
 It is important to take into account the long range
 interactions between pseudoparticles \cite{Polyakov}, so the vacuum of
 the theory is a Coulomb plasma of monopoles and antimonopoles, globally
 neutral and dilute at weak coupling. The monopole gas is conveniently
 described by a scalar field theory with cosine interaction, the coupling
 being proportional to the mean monopole density \cite{Polyakov}. The
 scalar field is dual to the photon in the sense of the usual
 electric--magnetic duality -- the monopoles are the sources of this
 field.

 When a probe charge is inserted in the vacuum, its electric field is
 screened by monopoles and form a tube with thickness of order of the
 correlation length in the Coulomb plasma. The surface spanned by the
 trajectory of the charge serves as a source of the dual scalar
 field.  The action of the corresponding classical configuration is
 proportional to the area of this surface \cite{Polyakov}. The
 surface appearing in the semiclassical calculations, therefore, can
 be interpreted as a world sheet of the string which confine the
 charges.  More recently, some progress have been made in the
 dynamical consideration of such strings \cite{Polyakov96}. The
 arguments were given \cite{Polyakov96} that beyond the semiclassical
 approximation string world sheet fluctuates and the Wilson loop
 average $W(C)$ in the gauge theory can be represented as a sum over
 surfaces bounded by the contour $C$ whose Boltzmann weight is
 determined by some string action.

 The three--dimensional adjoint Higgs model possesses interesting
 thermodynamic properties. It undergoes
 a deconfinement phase transition and in the high temperature phase
 linear forces between static charges are replaced by the Coulomb
 logarithmic interaction.  The universality arguments, as well as the
 renormalization group methods, were used to study this phase
 transition in a closely related model of lattice $U(1)$ gauge theory
 \cite{SY}.  The phase transition was shown to be of the
 Berezinskii--Kosterlitz--Thouless (BKT) type.  The reasons are based on
 the dimensional reduction of the monopole plasma at finite temperature
 to a two--dimensional Coulomb gas which is known to undergo the BKT
 phase transition \cite{KT}. Because at weak coupling the effects of the
 monopoles are exponentially small, the confinement scale of the theory
 is very large and the dimensional reduction should work well even for
 rather low temperatures. Thus, the deconfinement phase transition
 can be accurately described within the two--dimensional theory.

 In the present paper 3d adjoint Higgs model at finite temperature is
 studied in more detail.  We shall be primarily interested in the
 behavior of the Polyakov loops which measure the free energy of
 static charged sources and play the role of an order parameter for
 the deconfinement phase transition \cite{Svetitsky}. We find the
 operators corresponding to them in the effective sine--Gordon model.
 In two dimensions, the sine--Gordon theory is completely integrable
 and many quantities in it can be calculated exactly. The dimensional
 reduction enables us to utilize some of these exact results.

 The thermodynamics of 3d adjoint Higgs model is interesting by
 itself, but there exist some other motivations to study it. The
 point is that the dimensional reduction is expected to be a good
 approximation at comparably low temperatures. This fact allows us to use
 it in the study of the confinement phase. The theory considerably
 simplifies after passing from three to two dimensions and some principal
 questions become more tractable, in particular,
 the problem of string representation for Wilson loop averages.  The
 temporal degrees of freedom of the confining string decouple under
 the dimensional reduction and a sum over surfaces reduces to a sum
 over paths which is more familiar in the field theory and can be
 investigated in more detail. Another important question which can be
 studied with the help of the dimensional reduction concerns a
 spectrum of light degrees of freedom in the confinement phase.

 We also discuss the non--Abelian gauge symmetry restoration at high
 temperature \cite{KKL}. This permits to examine the phase diagram of
 the model more completely. An extrapolation of the results obtained
 leads to interesting predictions about the phase structure in
 the nonperturbative strong coupling region.

\section{3d adjoint Higgs model}

 Before considering the thermodynamics we briefly describe the
 properties of the theory involved at zero temperature following
 Ref.~\cite{Polyakov}.  The Euclidean action of the model has the form
\begin{equation}\label{act}
S=\int d^3x\,\left[\frac{1}{4g^2}\,\F_{\mu \nu }^a\F_{\mu \nu }^a
 +\frac{1}{2}\,D_\mu \Phi ^aD_\mu \Phi ^a
 +\frac{1}{4}\lambda \left(\Phi ^a\Phi ^a-\eta ^2\right)^2\right].
 \end{equation}
 The scalar field transforms in the adjoint representation. In this paper
 we consider $SU(2)$ gauge group, but a generalization to $SU(N)$
 with arbitrary $N$ is also possible \cite{DW,Snyderman}.

 The non--Abelian symmetry of \rf{act} is spontaneously broken
 to $U(1)$, unless $\eta ^2$ is not too small, when the transition to
 the symmetric phase can occur. This transition  was recently studied
 numerically in much detail \cite{HPST96,KLRS97}.
 The perturbative spectrum of the
 model consists of the massless photon, W and Higgs bosons with masses
 \begin{equation}\label{masses}
 m_W^2=g^2\eta ^2,~~~~~m_H^2=2\lambda \eta ^2,
 \end{equation}
 respectively.  In the perturbative regime,
 \begin{equation}\label{cond}
 m_H\sim m_W\gg g^2,
 \end{equation}
 the massive fields decouple at low energies and we are left with free
 $U(1)$ gauge theory. As it was shown in \cite{Polyakov}, this simple
 picture is spoiled by nonperturbative effects related to monopoles.

 The monopole solutions have the form
 \begin{mathletters}
 \label{monopoles}
 \begin{eqnarray}
 &&\A_\mu ^a=\varepsilon _{a\mu \nu }\,\frac{x_\nu
 }{r^2}\,\Bigl(1-f(r)\Bigr),
 \\
 &&\Phi ^a=q\eta \,\frac{x_a}{r}\,\Bigl(1-u(r)\Bigr),
 \end{eqnarray}
 \end{mathletters}
 where $r=|x|$ and $q=\pm 1$ -- is a magnetic charge. The functions $f$
 and $u$ fall exponentially at the distances of order $m_W^{-1}$
 or $m_H^{-1}$. At the origin they
 behave so that the solution is nonsingular.
 The monopoles have a
 finite action and their contribution can be calculated by the
 conventional semiclassical techniques. In the dilute gas
 approximation the Boltzmann weight of a single pseudoparticle and
 their interactions are treated separately.  The one--particle
 partition function $\zeta$ is then obtained by the loop expansion
 around the classical solution \rf{monopoles}:
 \begin{equation}\label{zeta} \zeta =
 \const\,\,\frac{m_W^{7/2}}{g}\,\e^{-\frac{4\pi
 m_W}{g^2}\,\epsilon(m_H/m_W)}.
 \end{equation}
 The exponential is the classical action of the monopole.
 The dimensionless function $\epsilon(m_H/m_W)$ varies from
 $\epsilon(0)=1$ \cite{BPS} to $\epsilon(\infty )=1.787...$ \cite{KZ}.
 The constant in the pre--exponential factor is determined by the
 loop corrections and is expected to be of order unity for the values of
 the parameters satisfying \rf{cond}.  However, it is known that the
 one--loop contribution diverges in the BPS limit $m_H\rightarrow 0$
 \cite{KS}.  Hence, the BPS limit lies outside the region of
 applicability of the semiclassical approximation.

 The monopoles interact as Coulomb charges of the magnitude
 $\sqrt{4\pi}/g$. The vacuum of the theory is, therefore, a Coulomb
 gas of monopoles and antimonopoles. At weak coupling the monopole gas is
 dilute. The dependence of the monopole density on couplings was studied
 numerically in Ref.~\cite{HPST96} and was found to be actually small
 in the perturbative regime.  In the dilute gas approximation, the
 monopoles contribute to the correlation functions via their classical
 long range fields.  Obviously, only the Abelian fields of the solution
 \rf{monopoles} survive on the distances much larger than the monopole
 size.  This can be checked by transforming the classical solution to the
 unitary gauge $\Phi ^1=0=\Phi ^2$. The remaining long range component
 $A_\mu \equiv \A_\mu^3$ obeys the superposition principle for
 multimonopole configurations.  Zero mode integration in the functional
 integral leads to the averaging over all configurations of this type.

 Although the contribution of the pseudoparticles is exponentially small,
 it has more important consequences than power--like perturbative
 corrections, and the monopoles should be retained in the low energy
 Abelian theory. Due to
 the Debye screening by monopoles the photon acquires the mass
 \cite{Polyakov}
 \begin{equation}\label{phmass}
 m_\gamma^2 =\frac{32\pi ^2\zeta }{g^2}
 \end{equation}
 and the Wilson loop expectation values exhibit an area law behavior
 with the string tension \cite{Polyakov,Snyderman}:
 \begin{equation}\label{stt}
 \sigma_0 =\frac{g^2m_\gamma }{2\pi ^2}.
 \end{equation}

\section{Partition function} \label{secpf}

 The partition
 function of the system defined by the action \rf{act} at the
 temperature $T$ is conventionally represented by the functional
 integral with periodic boundary conditions in the imaginary time:
 \begin{equation}\label{part}
 Z=\int [d\A][d\Phi ]\,\e^{-S-S_{\rm gf}-S_{\rm gh}},
 \end{equation}
 where $S_{\rm gf}$ and $S_{\rm gh}$ are gauge fixing and ghost terms,
 respectively. All fields are periodic in $x_0$ with the period $\beta
 =1/T$. The integral over $x_0$ in \rf{act} is also
 assumed to range from $0$ to $\beta $.

 At sufficiently low temperatures,
 \begin{equation}\label{condt}
 T\ll m_W,
 \end{equation}
 only Abelian degrees of freedom are relevant. Apart from the free
 photons, we must also take into account the monopole contribution. The
 corresponding classical field configurations now do not coincide
 with \rf{monopoles}, because they should respect the periodic boundary
 conditions.
 This can be easily achieved by considering the
 periodic chains of monopoles placed at the points with coordinates
 $x_\mu ^{(n)}=x_\mu +\delta _{\mu \,0}\,n\beta $.
 Since $\beta m_W\gg 1$,
 the distance between neighboring monopoles in the chain is much
 larger than their size and such classical configurations can be
 treated within the dilute gas approximation, so the one--particle
 partition function for the periodic monopole in this approximation
 is the same as in \rf{zeta}.  To be more precise, one should also
 take into account the Coulomb repulsion of elementary monopole from
 its images in the chain, but it is convenient to consider this
 repulsion as a part of the interaction energy, when we take into
 account the multimonopole configurations. For the interaction energy
 in the gas of the monopoles, therefore, we have:
 \begin{equation}\label{interaction}
 S_{\rm int}=\frac{2\pi }{g^2}\sum_{a,b}{\sum_{n}}'
 \frac{q_aq_b}{|\,x_a-x_b^{(n)}|},
 \end{equation}
 where the first sum runs over all pseudoparticles, $q_a=\pm 1$ are their
 magnetic charges and the prime means that the term
 with $n=0$ is omitted for $a=b$.

 The interaction energy is divergent in the infrared, unless the total
 magnetic charge, $\sum q_a$, is equal to zero. So, strictly speaking, it
 is necessary to insert the delta function $\delta _{\sum q_a,\,0}$ in the
 summation over all monopole configurations.  However, the neutrality
 condition in the Coulomb plasma is satisfied
 automatically \cite{Polyakov}, infrared divergencies cancel by
 themselves and there is no need to worry about them. These properties
 are essentially the consequences of the Debye screening.

 The properly regularized sum
 \begin{equation}\label{perpr}
 G(x )=\sum_{n=-\infty }^{+\infty }
 \frac{1}{\sqrt{{\bf x}^2+(x_0-n\beta )^2}}
 \end{equation}
 defines the periodic Green function of the Laplace operator:
 \begin{equation}\label{lap}
 -\partial ^2G(x )=4\pi \delta ({\bf x})\sum_{n=-\infty }^{+\infty }
 \delta (x_0-n\beta ).
 \end{equation}
 Therefore, in the dilute gas approximation, the partition function
 has the following form:
 \begin{equation}\label{dga}
 Z=Z_{\rm ph}
 \sum_{N=0}^{\infty }\frac{\zeta ^N}{N!}\sum_{q_a=\pm 1}
 \int \prod_{a=1}^{N}d^3x_a\,\e^{-\frac{2\pi }{g^2}
 \sum\limits_{a,b}q_aq_bG(x_a-x_b )}.
 \end{equation}
 The Boltzmann factor corresponds to the interaction of monopoles and
 antimonopoles, and the fugacity is determined by the one--monopole
 partition function.
 We denote a free photon contribution by $Z_{\rm ph}$:
 \begin{equation}\label{freeph}
 Z_{\rm ph}=\int [DA]\,\e^{-\frac{1}{4g^2}\int_{0}^{\beta }dx_0
 \int d^2x\,F_{\mu \nu }F_{\mu \nu } }.
 \end{equation}
 Here $F_{\mu \nu }$ is the Abelian field strength:  $F_{\mu \nu
 }=\partial _\mu A_\nu -\partial _\nu A_\mu $. We use the
 unitary gauge: $\Phi ^1=0=\Phi ^2$, in which $A_\mu\equiv\A^3_\mu  $.
 The correlation functions receive the contribution both from the free
 photon part and from the monopoles.

 At low temperature the partition function \rf{dga} describes the
 globally neutral Coulomb plasma. But, as the
 temperature is raised, the monopoles form bound states with
 antimonopoles and the system passes to the molecular phase. The
 existence of the BKT phase transition from the plasma to the
 molecular phase can be demonstrated by the following simple argument
 due to Kosterlitz and Thouless \cite{KT}.  The Green function
 \rf{perpr} behaves at large spatial separation, $|\bx|\gg\beta $, as
 \begin{equation}\label{ladis}
 G(x )\simeq-\frac{2}{\beta }\ln\left(|\bx|\mu \right),
 \end{equation}
 where $\mu $ is an IR cutoff. Restricting ourselves to the two--particle
 partition function, we find that the mean squared separation between
 monopole and antimonopole diverges at low temperature as
 \[
 \left\langle r^2\right\rangle\sim\int
 d^2x\,|\bx|^2\e^{\frac{4\pi }{g^2}\,G(x_0,\bx )}\sim
 \int^{\infty }dr\,r^{3-\frac{8\pi }{g^2\beta }}.
 \]
 But beyond the critical point,
 \begin{equation}\label{tc}
 T_c=\frac{g^2}{2\pi },
 \end{equation}
 the integral converges at large distances. Consequently, the mean
 separation becomes finite and monopoles and antimonopoles form the bound
 states.

 There is no Debye screening in the molecular phase of the monopole gas.
 On the other hand, the Debye screening of monopoles is responsible for
 the linear confining forces between electric charges. So, the BKT phase
 transition is associated with the deconfinement of electric charge
 in the adjoint Higgs model. The temperature dependence of the order
 parameter for the deconfinement phase transition -- the Polyakov
 loop -- is discussed in the next section.

\section{Polyakov loops}

 The Polyakov loop is a phase factor associated with the contour which
 closes due to the periodic boundary conditions:
 \begin{equation}\label{pl}
 L(\bx)=\e^{\frac{i}{2}\int_{0}^{\beta }dx_0\,A_0(x_0,\bx)}.
 \end{equation}
 It describes a static charge inserted in the vacuum at
 the point $\bx$. We consider charge 1/2
 Polyakov loops.  The reason is that they correspond to the matter field
 in the fundamental representation of $SU(2)$ -- after the symmetry
 breaking the latter splits into the two fields of charge $\pm 1/2$.

 The correlation functions of Polyakov loops play a distinguished role in
 gauge theories at finite temperature, since they measure the free energy
 of the static charge sources \cite{Svetitsky}.
 The expectation value of the Polyakov loop is equal to zero
 in the confinement phase, because the
 energy of a single charged particle is infinite.
 In principle, the Polyakov loop should acquire a
 non--zero expectation value in the deconfinement phase, but it is not
 the case for the model under consideration due to the infrared
 divergencies related to the low dimensionality of the problem \cite{SY}.
 More
 appropriate parameter  is the two--point correlator of
 the Polyakov loops, which is related to the interaction potential
 between particles of opposite charge \cite{Svetitsky}:
 \begin{equation}\label{intp}
 \left\langle L(\bx)L^{\dagger}(\by)\right\rangle=
 \e^{-\beta V(\bx-\by)}.
 \end{equation}

 In the confinement phase the potential grows linearly at large
 separation between charges, which
 is equivalent to the screening of the Polyakov loops:
 \begin{equation}\label{confin}
 \left\langle L(\bx)L^{\dagger}(\by)\right\rangle
 \sim\e^{-\beta\sigma|\bx-\by|},
 \end{equation}
 The screening length determines the string tension $\sigma $.
 The deconfinement transition is associated with the disappearance of
 the screening, and in the deconfinement phase the two--point
 correlator has a power--law behavior \cite{SY}:
 \begin{equation}\label{deconfin}
 \left\langle L(\bx)L^{\dagger}(\by)\right\rangle
 \sim|\bx-\by|^{-\beta \alpha},
 \end{equation}
 indicating the Coulomb logarithmic interaction of charged particles.

 In the dilute gas approximation, the monopoles contribute to the Polyakov
 loop via their classical fields.
 The long range field of the
 't~Hooft--Polyakov solution \rf{monopoles}
 in the unitary gauge coincides with that of a
 Dirac monopole:
 \begin{equation}\label{dmon}
 \tilde{F}_{\mu \nu }=q\varepsilon _{\mu \nu \lambda }\,\frac{x_\lambda
 }{r^3}.
 \end{equation}
 Here and below we mark by tilde the fields in the infinite Euclidean
 space which do not obey periodic boundary conditions. The gauge
 potentials for the Dirac monopole can be chosen as follows:
 \begin{mathletters}
 \label{dpo}
 \begin{eqnarray}
 &&\tilde{A}_0=-q\left(1-\frac{x_2}{r}\right)\frac{x_1}{x_0^2+x_1^2},
 \\
 &&\tilde{A}_1=q\left(1-\frac{x_2}{r}\right)\frac{x_0}{x_0^2+x_1^2},
 \\
 &&\tilde{A}_2=0.
 \end{eqnarray}
 \end{mathletters}
 The Dirac string in this gauge is directed along the $x_2$ axis.

 The relevant classical configurations at finite temperature are the
 periodic chains of monopoles. For them:
 \begin{equation}\label{pmon}
 A_\mu (x_0,\bx)=\sum_{n=-\infty }^{+\infty }\tilde{A}_\mu (x_0+n\beta
 ,\bx).
 \end{equation}
 To calculate the contribution of the periodic monopole to the Polyakov
 loop, we need not perform the summation in \rf{pmon} explicitly, since
 \[
 \int_{0}^{\beta }dx_0\,A_0(x_0,\bx)=\sum_{n=-\infty }^{+\infty }
 \int_{0}^{\beta }dx_0\,\tilde{A}_0 (x_0+n\beta,\bx)
 =\int_{-\infty }^{+\infty }dx_0\,\tilde{A}_0 (x_0,\bx).
 \]
 Substituting the field of the monopole \rf{dpo} for
 $\tilde{A}_0$ we find:
 \begin{equation}\label{intmon}
 \int_{0}^{\beta
 }dx_0\,A_0(x_0,\bx)=\left(2\arctan\frac{x_2}{x_1}\,-\pi\right)q .
 \end{equation}
\begin{figure}[t]
  \centerline{\epsffile{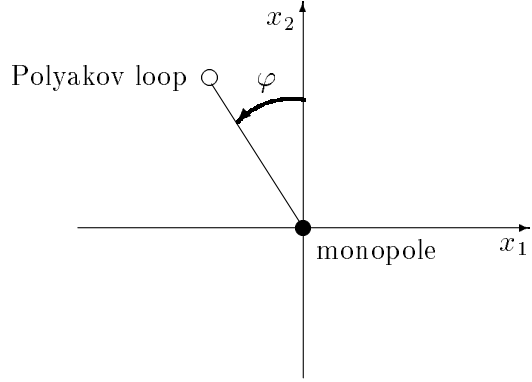}}
\caption {The definition of $\varphi(\protect\bx)$.}
\label{ugol}
\end{figure}
 This result shows that the contribution of a single monopole to the
 Polyakov loop $L(\bx)$ is equal to $\e^{iq\varphi(\bx) }$,
 where $\varphi(\bx)$ is an angle of the direction
 from the monopole position to the point $\bx$
 with $x_2$ axis, as depicted in Fig.~\ref{ugol}.
 The contributions
 of individual monopoles sum up and, for example,
 for the two--point correlator of
 Polyakov loops we get:
 \begin{eqnarray}\label{ll+}
 \left\langle L(\bx)L^{\dagger}(\by)\right\rangle
 =\frac{Z_{\rm ph}}{Z}
 \sum_{N=0}^{\infty }\frac{\zeta ^N}{N!}\sum_{q_a=\pm 1}
 \int \prod_{a=1}^{N}d^3x_a\,
 \exp&&\left[-\frac{2\pi }{g^2}
 \sum\limits_{a,b}q_aq_bG(x_a-x_b )
 +i\sum_{a}q_a\Bigl(\varphi _a(\bx)-\varphi _a(\by)\Bigr)
 \right.\non&&\left.
 \vphantom{
 -\frac{2\pi }{g^2}
 \sum\limits_{a,b}q_aq_bG(x_a-x_b )
 }
 -\frac{g^2}{16\pi }\int_{0}^{\beta }d\xi_0\,\int_{0}^{\beta }d\eta_0 \,
 \Bigl(G(\xi_0-\eta_0,{\bf 0} )-G(\xi_0-\eta_0,\bx-\by )\Bigr)\right].
 \end{eqnarray}
 The last term in the exponential represents a perturbative contribution
 of the free photons. The generalization of this formula to multipoint
 correlation functions is obvious.

 The partition function \rf{dga} has the convenient functional
 representation \cite{Polyakov}:
 \begin{equation}\label{sg3}
 Z=Z_{\rm ph}\int [d\chi ]\,\e^{-\int_{0}^{\beta }dx_0\int d^2x\,
 \left[\frac{g^2}{32\pi ^2}\,(\partial \chi )^2
 -2\zeta \cos\chi \right] }.
 \end{equation}
 The expansion of this path integral in $\zeta $ generates the
 grand canonical partition function for monopoles \rf{dga}, which can be
 checked using eq.~\rf{lap} in the calculation of Gaussian integrals over
 $\chi $.

 Correlation functions also admit the functional representation. Our
 present goal is to find the operators in the sine--Gordon theory \rf{sg3}
 corresponding to the Polyakov loops. The correlation functions which
 reduce to the form similar to \rf{ll+} have been considered
 in Ref.~\cite{JKKN} for 2d lattice model, closely related to the
 sine--Gordon theory.  Motivated by these results, we propose the
 following identification:
 \begin{mathletters}
 \label{pl'}
 \begin{eqnarray}
 &&L(\bx)=\e^{\frac{g^2}{8\pi }\int_{0}^{\beta }d\xi _0\,\int_{\bx}d\xi
 _i\,\varepsilon _{ij}\partial _j\chi },
 \\
 &&L^{\dagger}(\by)=\e^{\frac{g^2}{8\pi }\int_{0}^{\beta }d\xi
 _0\,\int^{\by}d\xi _i\,\varepsilon _{ij}\partial _j\chi }.
 \end{eqnarray}
 \end{mathletters}
 Here the contour of integration over $\xi _i$ goes to infinity in the
 $(x_1,x_2)$ plane or can end on the Polyakov loop of the opposite
 charge.  These formulas should be
 understood as operator equalities -- any correlation functions of the
 Polyakov loops in the $U(1)$ gauge theory with monopoles are equal to
 the correlation functions of the operators on the right hand
 side, where the averaging over $\chi $ is defined by the partition
 function \rf{sg3}.

 The identification \rf{pl'} is established by
 developing the correlation functions of the operators \rf{pl'}
 in $\zeta $. As a result, one recovers
 correlation functions for the Coulomb gas
 of type \rf{ll+}. In the specific case of
 the two--point correlator:
 \begin{equation}\label{ll+'}
 \left\langle
 \e^{\frac{g^2}{8\pi }\int_{0}^{\beta }d\xi _0\,\int_{\bx}^{\by}d\xi
 _i\,\varepsilon _{ij}\partial _j\chi }
 \right\rangle
 = \frac{Z_{\rm ph}}{Z}
 \sum_{N=0}^{\infty }\frac{\zeta ^N}{N!}
 \sum_{q_a=\pm 1}
 \int \prod_{a=1}^{N}d^3x_a\,
 \left\langle
 \e^{i\sum_{a}q_a\chi (x_a)
 +\frac{g^2}{8\pi }\int_{0}^{\beta }d\xi _0\,\int_{\bx}^{\by}d\xi
 _i\,\varepsilon _{ij}\partial _j\chi }
 \right\rangle_0,
 \end{equation}
 where $\left\langle\ldots\right\rangle_0$ denotes the Gaussian
 average over $\chi $. This average after some transformations can be
 represented in the following form:
 \begin{eqnarray}\label{zam}
 \left\langle
 \e^{i\sum_{a}q_a\chi (x_a)
 +\frac{g^2}{8\pi }\int_{0}^{\beta }d\xi _0\,\int_{\bx}^{\by}d\xi
 _i\,\varepsilon _{ij}\partial _j\chi }
 \right\rangle_0
 &=\exp&\left(
 -\frac{2\pi }{g^2}\sum\limits_{a,b}q_aq_bG(x_a-x_b )
 +\frac{i}{2}\sum_{a}q_a\int_{0}^{\beta }d\xi _0\,\int_{\bx}^{\by}
 d\xi _i\,\varepsilon _{ij}\partial _jG(\xi -x_a)
 \right.\non&&\left.
 \vphantom{
 -\frac{2\pi }{g^2}\sum\limits_{a,b}q_aq_bG(x_a-x_b )}
 -\frac{g^2}{32\pi}\int_{0}^{\beta }d\xi _0\,\int_{\bx}^{\by}d\xi
 _i\,\varepsilon _{ij}\int_{0}^{\beta }d\eta
 _0\,\int_{\bx}^{\by}d\eta _k\,\varepsilon _{kl}\partial _j\partial
 _lG(\xi -\eta )\right)
 \non
 &=\exp&\left[-\frac{2\pi }{g^2}
 \sum\limits_{a,b}q_aq_bG(x_a-x_b )
 +i\sum_{a}q_a\Bigl(\varphi _a(\bx)-\varphi _a(\by)\Bigr)
 \right.\non&&\left.
 \vphantom{
 -\frac{2\pi }{g^2}\sum\limits_{a,b}q_aq_bG(x_a-x_b )}
 -\frac{g^2}{16\pi }\int_{0}^{\beta }d\xi_0\,\int_{0}^{\beta }d\eta_0 \,
 \Bigl(G(\xi_0-\eta_0,{\bf 0} )-G(\xi_0-\eta_0,\bx-\by )\Bigr)\right].
 \end{eqnarray}
 This result is a simple generalization of the formula which have been
 used in Ref.~\cite{Zamolodchikov} to construct a path--integral
 version of soliton operators in 2d sine--Gordon theory. In the
 derivation we have assumed that the regularization is used in which
 $\delta (0)=0$.
 Substitution of this expression in equation~\rf{ll+'}
 reduces the latter to the form \rf{ll+}.

 The average
 \rf{ll+'} in the sine--Gordon theory reproduces not only the monopole
 contribution to the correlator of the Polyakov loops, but the whole
 answer containing also the free photon part. It is interesting that the
 same property holds for the string representation proposed in
 Ref.~\cite{Polyakov96} for the Wilson loop averages
 at zero temperature -- this representation
 automatically encounters the perturbative photon contribution. The
 Gaussian nature of the functional averages in each order of the
 expansion in $\zeta $ ensures the validity of the identification
 \rf{pl'} for arbitrary correlation functions. The operators \rf{pl'}
 possess a number of peculiar properties which are discussed in the next
 section.

\section{Dimensional reduction}

 The partition function \rf{sg3} defines the interacting
 field theory in three dimensions. Although the nonlinearity is caused
 by instantons and the coefficient before cosine, $\zeta $, is of order
 $\exp(-\const\,m_W/g^2)$, the interaction in the theory is not weak.
 The point is that $\zeta $ is the dimensional parameter and the
 strength of interaction depends on the scale. At very large
 distances, of order of the inverse photon mass $1/m_\gamma $ defined by
 eq.~\rf{phmass}, the effects of interaction are not small and they
 can not be neglected.  The consideration of a system at the
 finite temperature, actually, introduces a characteristic scale
 $\beta =1/T$.  Some simplifications do occur, when this scale is
 much smaller than the correlation length, i.e.\   when the
 temperature is not extremely low:  \begin{equation}\label{condt'}
 m_\gamma \ll T.  \end{equation} As before, the interaction can not
 be completely neglected, rather some degrees of freedom become
 irrelevant. The fields with nontrivial dependence on the time
 coordinate $x_0$ decouple, since their Matsubara frequencies are
 large compared to the characteristic mass scale determined by
 $m_\gamma $.  This argument justifies the dimensional reduction
 procedure, typical for the field theory at finite temperature.  As a
 result of this procedure, we are left with the two--dimensional
 sine--Gordon model:  \begin{equation}\label{sg2} S_{\rm
 reduced}=\int d^2x\,\left[\frac{g^2\beta }{32\pi ^2}\,(\partial \chi
 )^2 -2\zeta \beta \cos\chi \right].  \end{equation}

 In the Coulomb gas picture, the dimensional reduction can be understood
 as a substitution of the large distance asymptotics \rf{ladis} for
 the exact three--dimensional Green function in \rf{dga}. The
 discussion at the end of Sec.~\ref{secpf} demonstrates that the time
 dependence is irrelevant for the description of the deconfinement
 phase transition \cite{SY}. Moreover, the condition \rf{condt'} shows
 that the dimensional reduction is a good approximation in a wide range
 of temperatures below the phase transition. This circumstance allows us
 to use the reduced theory in the study of the confinement phase.  The
 dimensional reduction actually leads to substantial simplification,
 since 2d sine--Gordon model is a classic example of completely
 integrable field theory and some of its properties are known exactly.

 The conventional sine--Gordon action  is obtained after the rescaling
 of the field $\phi(\bx) \equiv\chi(\bx)\, g\sqrt{\beta} /4\pi $:
 \begin{equation}\label{sg}
 S_{\rm SG}=\int d^2x\,\left[\frac{1}{2}\,(\partial
 \phi )^2 -2\mu  \cos b\phi \right],
 \end{equation}
 where
 \begin{equation}\label{defb}
 b^2=\frac{16\pi^2 T}{g^2}=8\pi \,\frac{T}{T_c}\,.
 \end{equation}
 This parameter plays the role of a Plank constant in the sine--Gordon
 theory \cite{semicl}. Consequently, the low--temperature limit
 of the theory is semiclassical.

 The parameter $\mu $ depends on the renormalization scheme, i.e.\ on
 a definition of the operator $\cos b\phi $. The conventional
 regularization procedure is based on the normal ordering. Since the
 dimension of the operator $\no{\cos b\phi}$ is  $b^2/4\pi $, the
 coupling $\mu $ differs from its ``bare'' value $\zeta \beta $ by the
 factor $\Lambda ^{-\frac{b^2}{4\pi }}$, where $\Lambda $ is an UV
 cutoff.  It is worth mentioning that the dimensional arguments
 clarify the origin of the BKT phase transition -- at the critical
 point the dimension of the operator $\no{\cos b\phi}$ is 2, and the
 perturbation in \rf{sg} becomes marginal.

 The temperature defines the natural UV cutoff of 2d theory.
 Actually, the dimensional reduction is valid only at
 distances much larger than the inverse temperature, on smaller scales
 the dynamics is governed by the full 3d theory. In particular,
 the exact propagator $G(x)$ can not be replaced by its large
 distance asymptotics for $|\bx|\sim\beta $. However, due to the
 renormalizability all the effects of the dynamics at short distances
 can be absorbed into a multiplicative renormalization of $\mu $.
 Thus, the UV cutoff is proportional to the temperature with some
 numerical coefficient depending on the renormalization scheme.  For
 the normal ordering prescription this coefficient is calculated in
 Appendix:  \begin{equation}\label{tandl} \Lambda =\frac{\e^\gamma
 }{2}\,T\,, \end{equation} where $\gamma=0.5772\ldots$ is the Euler
 constant.  Thus, the renormalized value of $\mu $ is given by
 \begin{equation}\label{mur} \mu =\frac{\zeta }{T} \left(\frac{\e^\gamma
 }{2}\,T\right)^{-\frac{b^2}{4\pi }} =\frac{g^2m_\gamma^2 }{32\pi ^2T}
 \left(\frac{\e^\gamma }{2}\,T\right)^{-\frac{4\pi T}{g^2}}.
 \end{equation}

 \subsection{Polyakov loops and solitons}

 The peculiar property of the sine--Gordon theory is the presence
 of solitons in the spectrum of physical excitations. Classically,
 solitons are finite energy solutions of the equations of motion. They
 are very massive in the semiclassical region, but all more light
 particles can be viewed as soliton--antisoliton bound states
 \cite{semicl}. The existence of solitons is closely related to the
 hidden $U(1)$ symmetry generated by the topological current
 \begin{equation}\label{cur} j_i =\frac{b}{2\pi }\,\varepsilon
 _{ij}\partial _j \phi , \end{equation} which is identically conserved.
 Soliton carry a unit charge corresponding to this current.

 According to the results of the previous section, we associate with the
 Polyakov loops the following operators:
 \begin{mathletters}
 \label{pl2}
 \begin{eqnarray}
 &&L(\bx)=\e^{\frac{g^2}{8\pi T}\int_{\bx}d\xi
 _i\,\varepsilon _{ij}\partial _j\chi }
 =\e^{\frac{2\pi }{b}\int_{\bx}d\xi
 _i\,\varepsilon _{ij}\partial _j\phi },
 \\
 &&L^{\dagger}(\by)=\e^{\frac{g^2}{8\pi T}
 \int^{\by}d\xi _i\,\varepsilon _{ij}\partial _j\chi }
 =\e^{\frac{2\pi }{b}
 \int^{\by}d\xi _i\,\varepsilon _{ij}\partial _j\phi }.
 \end{eqnarray}
 \end{mathletters}
 It turns out that these operators have topological charges $1$ and $-1$,
 respectively. There are many possibilities to demonstrate this fact. For
 example, one can use OPE of the topological currents
 \begin{equation}\label{opejj}
 j_i(\bx)j_k(0)=-\frac{b^2}{8\pi ^3}\left(\delta
 _{ik}-2\,\frac{x_ix_k}{|\bx|^2}\right)\frac{1}{|\bx|^2}+\ldots
 \end{equation}
 to show that
 \begin{mathletters}
 \label{opejl}
 \begin{eqnarray}
 &&j_i(\bx)L(0)=-\frac{1}{2\pi }\,\frac{x_i}{|\bx|^2}\,L(0)+\ldots\,,
 \\
 &&j_i(\bx)L^{\dagger}(0)=\frac{1}{2\pi
 }\,\frac{x_i}{|\bx|^2}\,L^{\dagger}(0)+\ldots\,.
 \end{eqnarray}
 \end{mathletters}
 Defining the topological charge operator
 \begin{equation}\label{topchargeop}
 Q=\oint_C dx_i\,\varepsilon _{ik}j_k(\bx),
 \end{equation}
 where the contour $C$ encircles counterclockwise the origin, we
 readily find that $L$ ($L^{\dagger}$) behaves as charge $1$ ($-1$)
 operator under $U(1)$ transformations generated by $Q$:
 \begin{mathletters}
 \label{c1}
 \begin{eqnarray}
 &&Q\,L(0)=L(0), \\ &&Q\,L^{\dagger}(0)=-L^{\dagger}(0).
 \end{eqnarray}
 \end{mathletters}
 On the other hand, Polyakov loop, by its definition, creates an electric
 charge. This means that the following identification holds: ${\rm
 topological~charge}=2\times {\rm electric~charge}$, and the $U(1)$
 symmetry of the sine--Gordon model corresponds to the invariance of the
 original three--dimensional theory under global gauge
 transformations.

 The fact that Polyakov loops create solitons has important
 consequences. In particular, the large distance behavior of the
 two--point correlator of the Polyakov loops is governed by the
 lightest state with the topological charge one, i.e.\ by the
 soliton:
 \begin{equation}\label{llsol}
 \left\langle L(\bx)L^{\dagger}(\by)\right\rangle
 \sim\e^{-M|\bx-\by|}.
 \end{equation}
 Here $M$ is a soliton mass.
 Comparing this expression with \rf{confin} we find that the
 temperature dependence of the string tension is determined by the
 mass of the soliton:
 \begin{equation}\label{strten}
 \sigma (T)=TM.
 \end{equation}
 In the zero temperature limit the semiclassical approximation is
 valid. The classical mass of the soliton is equal to $8\sqrt{2\mu
 }/b $ \cite{semicl}, and we obtain:
 \[ \sigma (0)=\frac{8T\sqrt{2\mu }}{b}=\sqrt{\frac{8g^2\zeta }{\pi ^2}}
 =\sigma _0,\]
 where $\sigma _0$ is given by eq.~\rf{stt}. Thus, the result of
 Ref.~\cite{Snyderman} is recovered.

 In fact, the soliton mass in the sine--Gordon theory is known exactly
 \cite{smass}:
 \begin{equation}
 M=\frac{2\,\Gamma \left(\frac{p}{2}\right)}{\sqrt{\pi }\,\Gamma
 \left(\frac{p+1}{2}\right)}
 \left(\frac{\pi\, \Gamma \left(\frac{1}{p+1}\right)}{\Gamma
 \left(\frac{p}{p+1}\right)}\,\mu \right)^{\frac{p+1}{2}},
 \end{equation}
 where
 \begin{equation}\label{exmass}
 p=\frac{b^2}{8\pi -b^2}=\frac{T}{T_c-T}\,.
 \end{equation}
 After some algebra we obtain for the string tension:
 \begin{equation}\label{exten}
 \sigma(T) =\sigma _0\left(\frac{m_\gamma }{\e^\gamma \,
 T}\right)^{\frac{T}{T_c-T}}
 \left(\frac{\Gamma \left(\frac{T_c-T}{T_c}\right)}{\Gamma
 \left(\frac{T_c+T}{T_c}\right)}\right)^{\frac{T_c}{2T_c-2T}}\,
 \frac{\Gamma^2 \left(\frac{2T_c-T}{2T_c-2T}\right)}{\Gamma
 \left(\frac{2T_c-T}{T_c-T}\right)}\,.
 \end{equation}
 It is worth mentioning that the dependence of the soliton mass on $\mu
 $, and, consequently, of the string tension on $m_\gamma $,
 follows from a dimensional consideration. The
 remaining factor, which is a function of $b^2$, or, equivalently, of
 the adimensional ratio $T/T_c$, can not be found by elementary
 methods.

 \subsection{Deconfinement phase transition}

\begin{figure}[t]
  \centerline{\epsffile{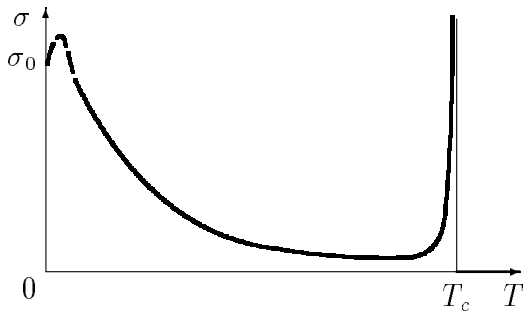}}
\caption {The string tension $\sigma (T)$.}
\label{ten}
\end{figure}

 The temperature dependence of the string tension is shown in
 Fig.~\ref{ten}.
 At low temperatures, $m_\gamma \ll T\ll T_c$, the string tension falls
 rapidly:
 \begin{equation}\label{lt}
 \sigma\simeq\sigma _0\left(\frac{m_\gamma }{\e^\gamma \,
 T}\right)^{\frac{T}{T_c}}\sim \sigma _0\exp\left(-\frac{4\pi
 ^2\epsilon(m_H/m_W)m_WT}{g^4}\right).
 \end{equation}
 The increase of $\sigma (T)$ at $T\sim m_\gamma $, certainly, is an
 artifact of the dimensional reduction which is inapplicable at such
 small temperatures. It is interesting that the low temperature
 behavior of the string tension is completely determined by the
 dimensional arguments.

 Let us turn to the critical behavior.  The string tension has an
 essential singularity at the point of the phase transition:
 \begin{equation}\label{ht}
 \sigma =\sigma _0\sqrt{\frac{\pi }{2}}\e^{\frac{3}{2}-\gamma }
 \left[\frac{m_\gamma ^2}{4\e^{2\gamma}T_c(T_c-T)
 }\right]^{\frac{T}{2T_c-2T}}\left(1+O\left(\frac{T_c-T}{T_c}\right)\right).
 \end{equation}
 The string tension decreases up to the
 temperature very closed to the transition point:
 \[
 T_m\simeq T_c-\frac{m_\gamma ^2}{4\e^{2\gamma -1}T_c},
 \]
 where it reaches vanishingly small value:
 \[
 \sigma _{\rm min}=\sigma _0\sqrt{\frac{\pi }{2}}\exp\left(2-\gamma
 -\frac{2\e^{2\gamma -1}T_c^2}{m_\gamma
 ^2}\right)\left(1+O\left(\frac{m_\gamma ^2}{T_c^2}\right)\right).
 \]
 Then the string tension grows rapidly and at the critical point turns to
 infinity. This result is essentially nonperturbative. It can not be
 obtained from the dimensional consideration. Were the dimensional
 analysis sufficient, the critical behavior of the string tension
 would be determined by the second factor in \rf{exten} and the string
 tension would vanish at $T=T_c$.  This is not the case. Actually, the
 third factor has a stronger singularity and becomes more important at
 the temperatures very closed to the point of the phase transition.

 \subsection{Confinement phase}

 Certainly, the reduced 2d theory is much simpler than the original 3d
 model.  On the other hand, the dimensional reduction is a good
 approximation up to very low temperatures, and one might think that the
 results obtained for the reduced model can be even extrapolated to zero
 temperature.  The fact that the string tension obtained from the mass of
 the soliton in two dimensions agrees with the result of the
 semiclassical analyses of Wilson loops directly in
 the three--dimensional theory \cite{Polyakov,Snyderman} supports this
 assumption. There are two points we are going to discuss:

 --- Perturbatively, the only light particle in the 3d adjoint
 Higgs model is the photon with the mass $m_\gamma $. Other degrees of
 freedom have substantially larger masses $m_W$ and $m_H$. It was
 conjectured that the dynamics in the intermediate region may be
 nontrivial and there are excitations with the masses $m_\gamma \ll m\ll
 m_W$ \cite{Polyakov96}. The known exact spectrum of the reduced theory
 provides a way to check this conjecture, since nonperturbative degrees
 of freedom should manifest themselves somehow at finite but
 small temperatures: $T_c\gg T\gg m_\gamma $.

 --- Another interesting problem is the string representation for the
 Polyakov loop correlators.

 We begin with the consideration of the spectrum. The semiclassical
 spectrum of 2d sine--Gordon theory \cite{semicl} turns out to be
 exact \cite{exact} and consists of solitons, antisolitons and their
 bound states (breathers). The lightest bound state can be identified
 with the perturbative excitation of the sine--Gordon field
 \cite{semicl} which, in turn, corresponds to a dual photon of the
 original 3d theory.

 Solitons and antisolitons
 disappear from the spectrum at $T\rightarrow 0$, since their mass
 determines the string tension by eq.~\rf{strten} and the string tension
 has the finite zero temperature limit. The exact masses of the breathers
 are given by the following expression \cite{semicl,exact}:
 \begin{equation}\label{bsm}
 m_n=2M\sin\frac{\pi np}{2}~ \left(n=1,2,\ldots\,;~n<\frac{1}{p}\right),
 \end{equation}
 where $M$ is the soliton mass and $p$ is defined in \rf{exmass}. The
 number of bound states depends on the coupling constant $b$, i.e.\  on
 the temperature. If the breathers survive the zero
 temperature limit, their number becomes infinite and they form a linear
 ``Regge trajectory'':
 \begin{equation}\label{bsmt=0}
 m_n\longrightarrow nm_\gamma .
 \end{equation}
 A possible interpretation of these particles is that they are
 threshold bound states of $n$ photons. The existence of excitations
 with masses which can be arbitrarily large compared to $m_\gamma $,
 but which are much smaller than $m_W$, gives the strong evidence of a
 nontrivial dynamics in the intermediate region.

 Probably the most important feature of $U(1)$ gauge theory with
 monopoles is its connection to string theory. On the level of classical
 equations of motion for the Wilson loop, a surface spanned by
 the loop parametrize naturally corresponding
 classical solution \cite{Polyakov}.
 The arguments were given that, beyond the semiclassical approximation,
 this surface fluctuates and Wilson loop average can be represented as a
 sum over random surfaces with some weight depending on the surface
 \cite{Polyakov96}.  This representation will be practically useful,
 if the effective string action has a simple form in some reasonable
 approximation.

 The same reasoning should be valid for the two--point correlator of
 Polyakov loops $\left\langle L(\bx)L^{\dagger}(\by)\right\rangle$. In
 this case, the string world--sheet is spanned by the contour
 $\L_{\bx}\Gamma _{\bx\by}\L_{\by}^{-1}\Gamma _{\bx\by}^{-1}$, where
 $\L_{\bz}=\left\{(x_0,\bz)|0\le x_0\le\beta \right\}$ and $\Gamma
 _{\bx\by}$ is some curve connecting $\bx$ with $\by$ in the
 $(x_1,x_2)$ plane.  We expect that at sufficiently high temperatures,
 $\sqrt{\sigma _0}\ll T\ll T_c$,  the temporal fluctuations of the string
 are irrelevant, in other words, the
 typical surfaces have the form $[0,\beta
 ]\times\Gamma _{\bx\by}$. So, the sum over random surfaces is replaced
 by a sum over paths:
 \begin{equation}\label{sumoverpath} \left\langle
 L(\bx)L^{\dagger}(\by)\right\rangle
 =\int_{\bX(0)=\bx}^{\bX(1)=\by}[d\bX]\,\e^{-S[\bX(\tau )]}.
 \end{equation}

 An important remark concerning this representation is in order.
 According to Ref.~\cite{Polyakov96}, the string world--sheet for a given
 field configuration $\chi (x)$ is defined by the equation $\cos\chi
 (X)=-1$. This is consistent with semiclassical analyses
 \cite{Polyakov96}. On the other hand, in the reduced 2d sine--Gordon
 model the same equation, $\cos b\phi (\bX)=-1 $, is a conventional
 definition of a soliton path \cite{Skyrme} for a field configuration
 $\phi (\bx)$.  Thus, the paths in the functional integral
 \rf{sumoverpath} are interpreted as soliton trajectories.
 It would be interesting to find such representation
 for the two--point correlator of the Polyakov loop operators
 \rf{pl2}.  Being the counterpart of the string representation for
 the Wilson loop averages, it may help to clarify the dynamics of the
 confining string.

 At present, we do not know how to construct the representation
 \rf{sumoverpath} with a reasonably simple action $S[\bX(\tau )]$.
 We know only one example of the simple sum--over--path
 representation for correlation functions -- the case when the
 operators are described by a local field theory. The
 sine--Gordon solitons are actually described by a local theory,
 namely, by
 the massive Thirring model \cite{Coleman,Mandelstam}:
 \begin{equation}\label{mt}
 S=\int d^2x\,\left(\bar{\psi }\gamma _\mu \partial _\mu \psi
 +M_T\bar{\psi }\psi +\frac{g_T}{2}\,\,
 \bar{\psi }\gamma _\mu \psi \, \bar{\psi }\gamma _\mu \psi \right),
 \end{equation}
 where $g_T=\frac{4\pi ^2}{b^2}-\pi$.
 However, the solitons obey Fermi statistics and soliton
 operators \cite{Mandelstam} (more precisely, their Euclidean--space
 counterparts \cite{Zamolodchikov}) differ from the Polyakov loops
 \rf{pl2} by an additional factor:
 \begin{mathletters}
 \label{so}
 \begin{eqnarray}
 &&\psi (\bx)=\left(
 \begin{array}{c}
 \no{\e^{\frac{2\pi }{b}\int^{\bx}d\xi
 _i\,\varepsilon _{ij}\partial _j\phi+\frac{i}{2}\,b\phi (\bx) }}\\
 -i\no{\e^{\frac{2\pi }{b}\int^{\bx}d\xi
 _i\,\varepsilon _{ij}\partial _j\phi-\frac{i}{2}\,b\phi (\bx) }}
 \end{array}
 \right)\,,
 \\
 &&\bar{\psi} (\bx)=\left(
 \begin{array}{c}
 i\no{\e^{\frac{2\pi }{b}\int_{\bx}d\xi
 _i\,\varepsilon _{ij}\partial _j\phi+\frac{i}{2}\,b\phi (\bx) }}\\
 \no{\e^{\frac{2\pi }{b}\int_{\bx}d\xi
 _i\,\varepsilon _{ij}\partial _j\phi-\frac{i}{2}\,b\phi (\bx) }}
 \end{array}
 \right)\,.
 \end{eqnarray}
 \end{mathletters}
 These operators have an interesting interpretation in
 the gauge theory. The operator
 $\e^{\pm\frac{i}{2}\,b\phi}=\e^{\pm\frac{i}{2}\,\chi }$ creates a
 monopole with magnetic charge $\pm 1/2$. So, from this point of view, the
 soliton can be regarded as a ``dyon'' -- the
 superposition of magnetic and electric charges, both equal to $1/2$ in
 magnitude.

 The free fermion propagator has the well known sum--over--path
 representation with an additional integration over
 Grassmannian world--line variables \cite{Polyakov87}. The
 world--line action in this representation is supersymmetric,
 although the supersymmetry is broken by boundary conditions.   This
 allows us to speculate that a modification of the Wilson loop by
 particular monopole operators in 3d theory has a string
 representation with supersymmetric world--sheet action. The loop
 corrections due to the four--fermion term in \rf{mt} can be regarded
 as contact string interactions.  Unfortunately, in the zero
 temperature limit both the mass $M_T$ and the coupling constant
 $g_T$ becomes infinite.

\section{Symmetry restoration}

 In the previous sections we have dealt with the low energy Abelian
 theory completely disregarding the massive fields. The only remnant of
 the spontaneously broken non--Abelian gauge symmetry relevant for this
 consideration was the presence of magnetic monopoles with finite action.
 This approximation, valid at sufficiently low temperatures, becomes more
 and more worse as the temperature is raised, since the thermal
 fluctuations decrease the mass scale of the
 complete non--Abelian theory. Ultimately, W--bosons become
 massless and the non--Abelian symmetry is restored. In the
 perturbative region \rf{cond}, the critical temperature should be very
 high, because the zero temperature value of the Higgs boson vacuum
 average is large and thermal fluctuations which decrease it to zero
 should be sufficiently strong. In this section we study the symmetry
 restoring phase transition in the framework of thermal perturbation
 theory.

 We add to the action \rf{act} the gauge fixing term
 \begin{equation}\label{gf}
 S_{\rm gf}=\frac{1}{2\alpha g^2}\,\int d^3x\,\left(\partial _\mu
 \A_\mu ^a+\alpha g^2\varepsilon^{abc}v^b\Phi ^c\right)^2,
 \end{equation}
 where $\alpha $ is a gauge parameter.
 This is a conventional choice which is convenient because it cancels a
 mixing between \mbox{W--bosons} and the complex scalar field $\Phi
 ^{\pm}=(\Phi^1\pm i\Phi ^2)/\sqrt{2} $ coming from the covariant
 derivative of $\Phi $ squared.  We always assume
 that the vacuum value of the scalar field is directed along the third
 axis in the color space and denote it by $v$:   $v^a=\delta ^{a3}\eta
 $.  The ghost term in the chosen gauge is
 \begin{equation}\label{gh}
 S_{\rm gh}=\int d^3x\,\left[\partial _\mu \bar{c}^aD_\mu c^a
 +\alpha g^2\eta \left(\Phi ^3\bar{c}^ac^a-\Phi ^a\bar{c}^a
 c^3\right)\right].
 \end{equation}

\begin{figure}[t]
  \centerline{\epsffile{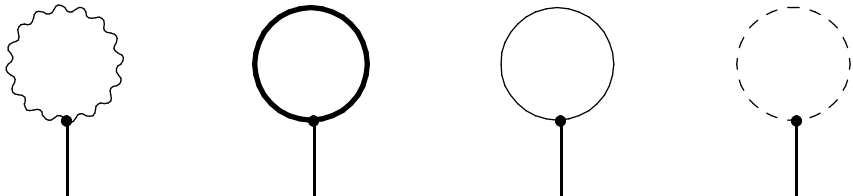}}
\caption {Tadpole diagrams corresponding to the
contribution of W--bosons, complex scalar field $\protect\Phi
^{\protect\pm}=(\protect\Phi ^1\protect\pm i\protect\Phi^2
)/\protect\sqrt{2}$, Higgs boson and ghosts to the vacuum expectation
value of the Higgs field.}
\label{tadpoles}
\end{figure}
 The reason for the symmetry restoration phase transition is the
 presence of the thermal corrections to the vacuum average
 $\left\langle\Phi ^3\right\rangle$ which is equal to $\eta $ at zero
 temperature. In order to
 find them, we expand $\Phi^3$ as $\Phi ^3(x)=\eta +\sigma (x)$
 and calculate $\left\langle\sigma \right\rangle$. To the lowest order of
 perturbation theory, $\left\langle\sigma \right\rangle$ is given by four
 tadpole diagrams depicted in fig.~\ref{tadpoles}. It is convenient to
 denote by $F(m^2)$ the contribution of a scalar loop:
 \begin{equation}\label{scl}
 F(m^2)=T\sum_{n}\int \frac{d^2p}{(2\pi )^2}\,
 \frac{1}{{\bf p}^2+\omega _n^2+m^2}-(T=0)
 ~\left(\omega _n=2\pi nT\right).
 \end{equation}
 The zero temperature part is subtracted because temperature independent
 contribution to $\left\langle\Phi ^3\right\rangle$ merely renormalize
 $\eta $. We can, actually, denote by $\eta $  the renormalized value of
 the Higgs condensate. Then the sum of the diagrams
 depicted in fig.~\ref{tadpoles} is found to be
 \begin{equation}\label{sumtadp}
 \left\langle\sigma \right\rangle=-\frac{1}{m_H^2}\left[
 2g^2\eta \left(2F(m_W^2)+\alpha F(\alpha m_W^2)\right)
 +2\lambda \eta F(\alpha m_W^2)+3\lambda \eta F(m_H^2)
 -2\alpha g^2\eta F(\alpha m_W^2)\right].
 \end{equation}

 The loop integral \rf{scl} is easily calculable by Poisson resummation
 \cite{Kapusta}:
 \begin{equation}\label{scl1}
 F(m^2)=\int \frac{d^2p}{(2\pi )^2}\,\,\frac{1}
 {\sqrt{{\bf p}^2+m^2}}
 \,\frac{1}{\e^ {\beta \sqrt{{\bf p}^2+m^2}}-1}
 =\frac{1}{2\pi \beta }\,\int_{m\beta }^{\infty }\frac{d\xi }{\e^\xi -1}.
 \end{equation}
 First of all, this formula shows that at
 the temperatures much smaller than the mass scale the
 corrections to the condensate are exponentially small.
 This result justifies the disregarding of the temperature corrections to
 the masses and the vacuum expectation value of the Higgs field in the
 previous sections. The corrections become comparable to
 $\eta $ at a very high temperature -- $T\gg m$. In this regime the
 function $F(m^2)$ can be expanded as
 \begin{equation}\label{scl2}
 F(m^2)=\frac{T}{4\pi }\,\ln\frac{T^2}{m^2}+O(m).
 \end{equation}
 Thus, we find with logarithmic accuracy:
 \begin{equation}\label{hc}
 \left\langle\Phi ^3\right\rangle=\eta +\left\langle\sigma
 \right\rangle
 \simeq\eta -\frac{T\eta} {4\pi m_H^2}
 \left(5\lambda +4g^2\right) \ln\frac{T^2}{m_W^2}.
 \end{equation}
 The dependence on $\alpha $ drops out from the final answer, as it
 should.

 At the point of the phase transition $\left\langle\Phi ^3\right\rangle$
 turns to zero. The critical temperature,
 \begin{equation}\label{t*}
 T_*=\frac{4\pi
 m_W^2}{g^2\left[\left(5+8\,\frac{m_W^2}{m_H^2}\right)\ln\frac{m_W}{g^2}
 +O\left(\ln\ln\frac{m_W}{g^2}\right)\right]},
 \end{equation}
 is very high in the perturbative region -- $T_*\gg m_W$, as expected.
 The symmetry restoration phase transition, most probably, is the
 second order one, since the condensate of the Higgs field
 is continuous at the critical point in the one--loop approximation.

\section{Discussion}

\begin{figure}[t]
  \centerline{\epsffile{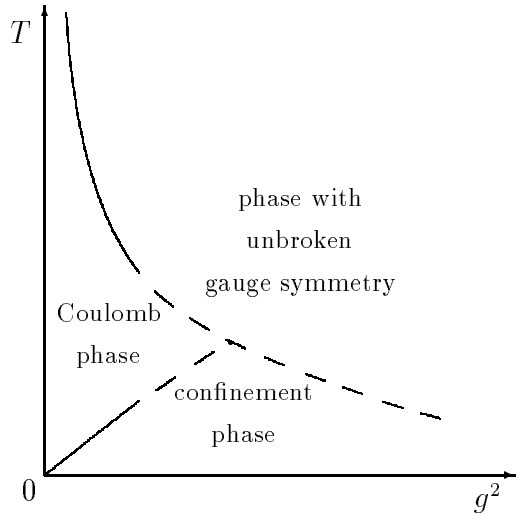}}
\caption {The phase diagram of 3d adjoint Higgs model.}
\label{faza}
\end{figure}

 The phase structure of 3d adjoint Higgs
 model can be studied in detail
 at weak coupling. The system
 undergoes the deconfinement transition of the BKT type and the second
 order phase transition associated with the restoration of the
 non--Abelian gauge symmetry. The temperatures of these transitions
 differ in  the order of magnitude: $T_*\gg T_c$. But $T_*$ rapidly
 decreases with the increase of the coupling $g^2$ at fixed $m_W$ and
 $m_H$, while $T_c$ grows. So, at $g^2\sim m_W$ the lines of
 deconfinement and symmetry restoration phase transitions meet at the
 triple point (Fig.~\ref{faza}).
 At sufficiently large $g^2$ the non--Abelian symmetry is restored
 directly in the confinement phase. The strong coupling behavior can not
 be described by perturbative methods and we have nothing to say about
 the phase transition between confinement and symmetric phases, but
 numerical results of Refs.~\cite{HPST96,KLRS97} show that at zero
 temperature this is a first order transition terminating at some point,
 so that symmetric and confinement phases are connected by analytical
 continuation. Probably, the same is true for the thermal phase
 transition as well.

 Our consideration of the critical behavior at the deconfinement
 phase transition is based on the operator representation for the
 Polyakov loops in the effective sine--Gordon model describing the
 Coulomb gas of the monopoles. After the dimensional reduction
 Polyakov loop operators turn out to be closely connected with the
 solitons in 2d sine--Gordon theory. Using the known exact value of
 the soliton mass \cite{smass} we found the string tension as a
 function of temperature, eq.~\rf{exten}. The string tension exhibits
 rather unusual behavior near the phase transition.  It grows rapidly
 in the vicinity of the transition point and at the critical
 temperature  diverges as $\exp(-\ln\Delta/2\Delta )$, where $\Delta
 =(T_c-T)/T_c$.

 The interpretation of the known facts about 2d sine--Gordon model
 enables us to draw some conclusions about the properties of the
 confinement phase of 3d adjoint Higgs model.  There are strong arguments
 in favor of nontrivial dynamics at the intermediate scales between
 $m_\gamma $ and $m_W$. Perhaps, the most interesting result is a
 possible appearance of strings with a world--sheet supersymmetry,
 but this is only a conjecture which requires more serious confirmations.

\acknowledgments

 The work of
 N.A.\ was supported in part by
 INTAS\- grant \mbox{94--2851} and
 RFFI grant \mbox{96--02--19184a}.
The work of K.Z.\ was supported in part by
 CRDF grant \mbox{96--RP1--253},
 INTAS\- grant \mbox{94--0840},
 RFFI grant \mbox{97--02--17927}
 and grant \mbox{96--15--96455} of the support of scientific schools.

 \appendix
 \section*{UV scale of the reduced theory
 \label{uvscale}}

 The normal ordered operator $\no{\cos b\phi }$ is defined by the
 two--point function in the free theory
 \begin{equation}\label{defnocos}
 \left\langle\no{\e^{ib\phi (\bx)}}\no{\e^{-ib\phi
 (\by)}}\right\rangle_0\equiv |\bx-\by|^{-\frac{b^2}{2\pi }}.
 \end{equation}
 On the other hand, eq.~\rf{sg3} determines
 the Gaussian average in three dimensions to be:
 \begin{equation}\label{defcos}
 \left\langle\e^{i\chi (x)}\e^{-i\chi (y)}\right\rangle_0
 =\e^{\frac{4\pi }{g^2}\bigl(G(0)-G(x-y)\bigr)},
 \end{equation}
 where $G(x)$ is defined by \rf{perpr}. The Coulomb gas representation
 of the monopole partition function \rf{dga} implies the particular
 regularization of the divergent quantity $G(0)$. It is defined by
 eq.~\rf{interaction}:
 \begin{equation}\label{regsum}
 G(0)-G(x-y)=\sum_{n\neq 0}\frac{1}{\beta |n|}-
 \sum_{n}\frac{1}{\sqrt{(\bx-\by)^2+(x_0-y_0-n\beta )^2}}
 =\frac{2}{\beta }\,\ln\frac{\e^\gamma |\bx-\by|}{2\beta }
 +O\left(\e^{-\frac{2\pi |\bx-\by|}{\beta }}\right).
 \end{equation}
 Here $\gamma $ is the Euler constant.
 The dimensional reduction is equivalent to the omitting of exponential
 terms. Hence, we obtain:
 \begin{equation}\label{regcos}
 \left\langle\e^{i\chi (\bx)}\e^{-i\chi (\by)}\right\rangle_0
 = \left(\frac{\e^\gamma}{2}\,T\right)^{-\frac{8\pi T}{g^2}}
  |\bx-\by|^{-\frac{8\pi T}{g^2}}
 \equiv\Lambda ^{-\frac{b^2}{2\pi }}|\bx-\by|^{-\frac{b^2}{2\pi }}.
 \end{equation}
 Comparing this equation with the correlation function of the normal
 ordered operators \rf{defnocos} we find that the definition of the
 operator $\cos b\phi $ in the monopole partition function is
 connected with the normal ordering prescription by the
 multiplicative renormalization:  \begin{equation}\label{regdefcos}
 \cos b\phi =\cos\chi =\Lambda^{-\frac{b^2}{4\pi }}\no{\cos b\phi },
 \end{equation}
 and the UV scale is given by
 \begin{equation}\label{lambdadef}
 \Lambda =\frac{\e^\gamma }{2}\,T\,.
 \end{equation}

\end{document}